\documentclass[conference]{IEEEtran}
\IEEEoverridecommandlockouts
 
\pdfoutput=1

\usepackage{tabulary}
\usepackage{times}
 
\usepackage[compact]{titlesec}
\usepackage[utf8]{inputenc}
\usepackage[hyphens]{url}
\usepackage[colorlinks=true,urlcolor=blue,citecolor=red]{hyperref}
\usepackage{wrapfig}
\usepackage{listings}
\usepackage{subfigure}
\usepackage{paralist}
\usepackage{graphicx}
\usepackage{amsmath}
\usepackage{amssymb}
\usepackage{comment}
\usepackage{enumitem}
\usepackage{xspace}
\usepackage{varwidth}
\usepackage{paralist}
\usepackage[most]{tcolorbox}
\usepackage{booktabs}
\usepackage{pgfplotstable}
\pgfplotsset{compat=1.17}
\usepackage{longtable}
\usepackage{caption}
\usetikzlibrary{positioning,shapes,arrows,trees}

\usepackage{soul}
\usepackage[most]{tcolorbox}

\usepackage{color,xspace,listings,comment}
 \newcommand{\name}{\textsc{FTTN}\xspace}

\usepackage{comment}
\usepackage{amsmath,amssymb,amsfonts}
\usepackage{algorithmic}
\usepackage{graphicx}
\usepackage{textcomp}
\usepackage{xcolor}

\usepackage{pifont}% http://ctan.org/pkg/pifont
\newcommand{\cmark}{\ding{51}}%
\newcommand{\xmark}{\ding{55}}%
 \usepackage{array}
\usepackage{multirow}
\usepackage[ruled, vlined, linesnumbered]{algorithm2e}%
\begin{document}

% \title{Microbenchmarking AMD GPU FP Math\\
% % {\footnotesize \textsuperscript{*}Note: Sub-titles are not captured in Xplore and
% % should not be used}
% \thanks{Identify applicable funding agency here. If none, delete this.}
% }
%\title{Discovery of  Floating-Point and   Performance Differences Between NVIDIA and AMD GPUs}

\title
{\name: Feature-Targeted Testing 
for Numerical Properties of  NVIDIA \& AMD Matrix Accelerators}

% Discovery of Floating-Point Differences Between NVIDIA and AMD GPUs: A Testing-Guided Approach

\author{\IEEEauthorblockN{Xinyi Li}
\IEEEauthorblockA{\textit{Kahlert School of Computing} \\
\textit{University of Utah}\\
 USA \\
xin\_yi.li@utah.edu}
\and
\IEEEauthorblockN{Ang Li}
\IEEEauthorblockA{\textit{Pacific Northwest National} \\
	\textit{ Laboratory}\\
	USA \\
	ang.li@pnnl.gov}
\and
\IEEEauthorblockN{Bo Fang}
\IEEEauthorblockA{\textit{Pacific Northwest National} \\
\textit{Laboratory}\\
USA \\
 bo.fang@pnnl.gov}
\and
\IEEEauthorblockN{Katarzyna Swirydowicz}
\IEEEauthorblockA{\textit{\hspace{5em}Pacific Northwest National\hspace{5em} } \\
\textit{Laboratory }\\
USA \\
kasia.swirydowicz@pnnl.gov}
\and
\IEEEauthorblockN{Ignacio Laguna}
\IEEEauthorblockA{\textit{Lawrence Livermore National} \\
\textit{Laboratory }\\
USA \\
ilaguna@llnl.gov}
\and
\IEEEauthorblockN{Ganesh Gopalakrishnan}
\IEEEauthorblockA{\textit{Kahlert School of Computing} \\
\textit{University of Utah}\\
USA\\
ganesh@cs.utah.edu}
}

\maketitle

\begin{abstract} 
 NVIDIA Tensor Cores and AMD Matrix Cores (together called Matrix Accelerators)
are of growing interest in high-performance computing and machine learning owing
to their   high  performance at low power consumption. Unfortunately, very few
facts are publicly documented about some of their attributes that can affect
answers computed on identical code. Examples of such  features are the number of
extra precision bits, accumulation order of addition, and  predictable subnormal
number handling   during computations. We demonstrate that the lack of
information on how these features differ across two matrix accelerators can
make it impossible to reliably port codes across GPUs containing these
differing accelerators. In response to this challenge, this paper offers a
collection of tests that are based on a precise understanding of the
IEEE floating-point standard as well as previously discovered
formal results about the impact of floating-point
features on numerical behavior. By running these tests on a large number of
widely used and  recent GPUs, we show that our tests can 
unearth feature differences that affect computed results.
We exhibit these differences across five floating-point formats,
four standard rounding modes and additional four feature combinations
including those relating to rounding and preservation of extra precision bits.
This extensive testing
demonstrates the versatility of our tests in picking up
salient differences that can affect numerical behavior across this space.
As further proof of the discriminative power of our approach,
we design a simple matrix-multiplication test with the matrix
entries designed with insights gathered from our feature-tests.
We executed this very simple test on five platforms, producing
different answers: V100, A100, and MI250X produced 0,
MI100 produced 255.875, and Hopper H100 produced 191.875.
There is no prior work that shows that a simple test like
this can produce three different answers on five different
platforms---raising concern that one carefully understand
Matrix Accelerator features before porting code across them.\end{abstract}


\section{Introduction}

\label{sec:intro}
We are in an era
of rising computing hardware heterogeneity where many new CPU and GPU components are introduced in rapid succession~\cite{rising-heterogeneity}, and
are fueling 
performance 
advances in HPC and ML: from drug discovery to climate simulations and beyond.
While no scientist
aims to achieve higher
performance  at the
expense of correctness,
   ensuring correctness has
become a serious 
challenge given the sheer number of hardware units and the rapidity of their adoption.
Specifically in the realm of GPU-based accelerators,
programmers are interested in 
testing codes developed for NVIDIA GPUs on
AMD GPUs that are becoming 
available: MI100 at first, MI250X now
and soon MI300 that
will be used in the upcoming El Capitan Exascale machine\footnote{\url{https://www.llnl.gov/article/49131/llnl-scientists-eagerly-anticipate-el-capitans-potential-impact}},
with earlier AMD models already in use
in Oakridge OLCF Frontier\footnote{\url{https://www.olcf.ornl.gov/olcf-resources/compute-systems/frontier/}}.
Unfortunately, documentation about many aspects of  these 
GPUs is found seriously lacking in terms of numerical aspects.
Unanswered questions not only pertain to particular behaviors such as precision loss for a specific operator but also {\bf important features} such as
the  rounding modes supported, fused-multiply-addition (FMA)
details, the number of extra precision bits held inside, the granularity of their block fused-multiply-add, etc.
The {\em presence or absence of} these features significantly influences numerous high-stakes algorithms in fields like HPC and ML.
At present, programmers resort to running many applications and monitoring the results; this process is not rigorous or scalable, and a new
program might one day cause a serious result difference.
It is highly desirable to have a set of straightforward tests that
can quickly pick up salient feature differences between GPUs, but these
do not exist.
Our primary contribution in this paper
is a rigorous methodology that has enabled us to create 
such discriminatory tests for 
NVIDIA and AMD GPUs, with the methodology generalizable and applicable
to future GPUs.

\paragraph{Matrix Accelerators}

While the general lack of information that affects reliable code porting across GPUs is 
well-acknowledged,
{\em matrix accelerators} pose an 
even higher degree of difficulty because of their growing importance and even more dearth of information---{\bf  hence forming the central
focus of this paper}.
We use the term ``Matrix Accelerator''
as a generic term to
refer to what
NVIDIA calls ``Tensor Cores~\cite{tensorcore}''
and
  AMD 
  calls   ``Matrix Cores~\cite{matrix-cores}.''
Matrix
Accelerators 
    are 
indispensable for achieving today's performance levels
in ML.
It is safe to say that 
language-level models
(e.g.
ChatGPT) will not have happened without Tensor Cores (ML training will take at least 10 times longer without the acceleration of Tensor Cores~\cite{narayanan2021efficient}).
Naturally, matrix accelerators have caught
the eye of 
HPC designers
who see its
4$\times$ speedup with
80\% less energy
consumption~\cite{blanchard:hal-02491076}
a real avenue toward much
faster and energy-efficient
codes as promoted in prominent articles~\cite{jack-royal}.
The work in this paper {\em is designed to help   programmers
  port code across matrix accelerators more reliably, based on the commonality of features that our tests help confirm.}

Unfortunately,   matrix accelerators are described in the literature mainly with respect to their usage and not numerical properties.
With the growing number of
these units in
upcoming GPUs~\cite{tpu},
  this situation
poses a serious impediment to
those wanting to use them
for HPC or port code across
two different models.
Additionally, designers
have to keep in mind 
 the variety of
 number formats supported
 by them, and variety of 
hardware features.

  What makes this variety 
troubling is that many of
these details (barring   a few basic
aspects of floating-point arithmetic supported) are  not documented anywhere
in a manner
that is easily accessible to the general public.
We show in this work that 
the undocumented feature 
differences can
affect computational result
portability in an
extreme manner.
%
% While one may argue that the
% actual sensitivity to changes in these parameters is 
%   application-dependent,
% even this argument must be backed 

In today's HPC software development approaches, not being able to use a new GPU or
its matrix accelerator as soon as they become available is
 a serious handicap: one cannot plan code migration, evaluate previous applications on  new machines, or provide timely  feedback to vendors.
Going 
``eye-ball result
agreements'' on test codes (often
today's approach) runs the
risk of the unexamined
cases.
The community urgently needs
approaches  that can reveal the
differences in terms of
critical numerical features that are
bound to affect some (future) program, thus
serving as early warning.
Such an approach applicable to CPUs or GPUs in general, and matrix accelerators 
in particular is the
key goal of this paper.

To provide some more evidence,
  recall that
if one uses 32-bit floating-point format (FP32), one can expect
a result-difference   in the 8th {\em decimal fraction} position; this is because
FP32's round-to-nearest
rounding can guarantee 7 (fractional) digits of accuracy.
The corresponding number
for the 16-bit format
(FP16, used internally by matrix accelerators) is three fractional digits being guaranteed.
However, if one ports an FP16 implementation of an HPC routine
across platforms and obtains a difference in the {\em third} decimal fraction position, 
then 
it is perhaps worth investigating what caused the
error to 
exceed what
precision arguments imply.

In fact, we could create 
a  focused test 
that went much further!
This test
performed
matrix multiplication  
based on our understanding of feature differences across matrix accelerators,
obtaining
  the following answers:
 V100, A100, and MI250X produced 0;
MI100 produced 255.875; and Hopper H100 produced 191.875.
This error is clearly far more serious: not merely the fourth {\em fractional} digit but the hundredth digit has been affected---an error that is
 {\em six orders of magnitude
 higher}.
Such results have never been demonstrated before, thus making 
the work in this paper---specifically our straightforward tests---important to put in the hands of today's programmers (we will release all our tests upon acceptance).

\paragraph{Problem Addressed}

The problem
addressed is
 the design
 of 
 a simple  and practical approach to  check whether critical feature differences lurk in execution units.
 The tests 
 ought to be based on straightforward
 logic pertaining to basic floating-point
 facts, allowing them to be easily extended
 to  newer model
 GPUs and Matrix Accelerators being developed 
 by multiple companies
 (e.g., Google TPUs~\cite{tpu} models; such accelerators
 may be of
 interest to HPC designers
 for their power/performance
advantages).
Running
larger
test programs 
such as actual large-scale numerical
solvers does not meet
this need, as
there is significant overhead associated with making 
programs run---especially
on new hardware
where libraries
and compiler
features might
be missing.

%
% For instance, one might tell a programmer that
% the arithmetic-logic-unit (ALU) is keeping only
% one extra-precision bit, and not the recommended
% 3 (namely {\it guard, rounding, sticky}).
%

Our feature-directed testing approach
goes to a considerable distance in 
meeting these idealized goals.
By showing the presence/absence of a feature,
it helps answer how whole classes of
programs can be affected, as will
be shown (\S\ref{sec:new-xl-section}).

\subsection{Related Work and Improvements Over Them}

Correctness of GPU numerical behavior
is a vast topic; given 
space restrictions, we study only those that target matrix accelerators or contrast different GPUs.
In~\cite{mikaitis2023monotonicity,higham-tensor-cores},
the concept of
{\em  monotonicity} (applicable to vector reductions) 
was studied using a testing-guided approach
(the mention of monotonicity as a desired property appears even earlier~\cite{nathalie}).
Monotonicity is true 
with respect to 
two vectors A and B
of equal length
if  
 the result of adding all elements of A
must not  exceeding that produced by all the
elements of B
 {\em provided}
$A$ 
is  position-wise
less than  or
equal to B
 (i.e., $A[i]\leq B[i]$).
Monotonicity 
 can be violated 
 if (1)~the reduction order is not externally
 controllable, and
 (2)~there aren't enough {\em extra precision bits} (specifically, 3 bits) provided within the arithmetic unit~\cite{mikaitis2023monotonicity}.
In this sense, their work set the stage for focusing on
feature differences; we advance this direction 
significantly in this paper.

The question of numerical portability 
between various GPUs
including AMD GPUs 
has been studied in~\cite{innocente-zimmerman}.
They target
math function results produced by 
earlier generation
GPUs  and libraries, with
no coverage of
matrix accelerators.
The paper~\cite{prior-amd} targets performance
portability  
across NVIDIA and AMD MI-100 GPUs
The paper~\cite{rydahl2023precision}   tests
and reports the
extent to which
 numercial errors exhibited
 by standard library
 math functions vary across platforms.

\begin{table*}[ht]
\centering
\caption{Floating-Point Format Comparison (TF32 is a proprietary format and can be implementation-dependent). The more the mantissa bits, the lesser the $ulp$. Notice how BF16 sacrifices mantissae for higher dynamic range (larger exponent size)}
\begin{tabular}{|c|c|c|c|c|c|c|}
\hline
\textbf{Format} & \textbf{Sign Bit (S)} & \textbf{Exponent Bits (E)} & \textbf{Mantissa Bits (F)} & \textbf{Min. Exponent ($e_{min}$)} & \textbf{Max. Exponent ($e_{max}$ )} & \textbf{\shortstack{$ulp$, i.e.\\ulp(1)}} \\
\hline
 {FP16} & 1 & 5 & 10 & -14 & 15 & $2^{-10}$\\
\hline
 {FP32} & 1 & 8 & 23 & -126 & 127 & $2^{-23}$  \\
\hline
 {FP64} & 1 & 11 & 52 & -1022 & 1023 & $2^{-52}$\\
\hline
 {BF16} & 1 & 8 & 7 & -63 & 63 & $2^{-7}$\\
\hline
 {TF32} & 1 & 8 & 10 & -126 & 127 & $2^{-10}$\\
\hline
\end{tabular}
\label{tab:floating_point_comparison}
\end{table*}

%\input{key-insights.tex}

% \subsection{Improvements over State of the art}

%\input{improvements-over-prior.tex}

\noindent{\bf Summary of Contributions:\/}

\begin{itemize}

\item We offer a novel set of
feature-targeted tests, with
 clear evidence that 
 if such features are not
 preserved across the source
 and target platform, 
 execution results may seriously
 differ.

 \item 
Our tests form a pipeline
with earlier tests confirming/denying
certain features and
later feature tests taking advantage
of it to
unambiguously confirm/deny a
second feature, and so on.

\item Our tests are simple enough to be run on early access machines that may not have full-fledged libraries or runtimes, yet powerful enough to confirm (or refute) the status of higher level features supported in hardware.

\item With the availability
of AMD machines, 
our work meets a critical need of moving code from NVIDIA to AMD for full comparisons.
We also point out the dangers
of doing porting in reverse,
when subnormal support is
missing
(this relates to support for
hardware trapping of exceptions).

\item This is the most extensive testing 
of matrix accelerators
to date that we are aware of, {\em including
the first results on H100 about its numeric features.}

\item Our tests indicate that AMD MI250X is closer to CPU behavior, thus perhaps requiring fewer changes during CPU-to-GPU porting.

    \end{itemize}

\noindent{\bf Roadmap:} We provide self-contained background crucial to understand the testing approaches in this paper (\S\ref{sec:bg}).
The main technical part of this
paper is
higher-level feature 
testing as 
applied to matrix 
accelerators (\S\ref{sec:matrix-accel-ma}).
The extent to which feature differences caused the results of a basic matrix multiplication routine to jump across three values on five GPUs is then presented (\S\ref{sec:new-xl-section}).
How our tests characterized three GPUs across five precision formats is   summarized in a
comprehensive results table (\S\ref{sec:ma-testing-big-table}).
 Conclusions follow
(\S\ref{sec:conc}).
%--

\section{Background}
\label{sec:bg}
Floating-point arithmetic is a vast domain, and our objective here is to provide a
{\em high level  overview} of facts crucial to understand how we designed our tests.

\subsection{Floating-point background}
% The subject of floating-point
% arithmetic is vast, and we only
% attempt an overview sufficient to
% appreciate the details presented in this
% paper.
% %
% The basic layout of various floating-point
% formats (FP16, FP32, FP64) are
% widely discussed in the
% literature (e.g., \cite{fp-handbook}).
% %
% Newer formats such as 
% brain float (BF~\cite{bfloat-16})
% and Tensor Float (TF~\cite{tf32}) are
% being proposed largely supporting ML.

% The IEEE standard
% defines
%      various rounding modes,
%      and widely cited work~\cite{goldberg1991every}
%      discuss the importance of
%      maintaining extra bits
%      including
%      introduce guard bit, round bit and sticky bits.\ggcmt{XINYI plz clean up and add 1-2 sentences for these.  
% Also
%     explain that rounding mode is dependent on the number of extra bits (guard, round and sticky).
% Throw in an FMA explanation - fused
% operation - also here.}

%
%

A floating-point
number~\cite{fp-handbook} $x=(s,e,m)$
consists of a single sign bit 
$s$, 
a mantissa (also called significand) $m$ (of 
23 bits)
representing a value in the real interval $(0,2)$
and an exponent $e$
(of 8 bits, typically presented as a biased integer).
Regard $m$ and $e$ as the intended (i.e., ignore the bias in $e$) real-numbered.
Then
the  value
of  the
floating-point number is  
$$\text{fp}\_\text{value}(x) = (-1)^{s} \cdot m \cdot 2^{e} $$ (see Table~\ref{tab:floating_point_comparison} for other pertinent details).
%
% x = (s,e,m)
% fpvalue(x)
% fpvalue(s,e,m)
%
Aiming for a unique and convenient representation, the mantissa $m$ remains in the range of $[1, 2)$ whenever $e > e_{min}$, and hence can be expressed as a fraction
$1.(..23 bits..)$
which is called the {\em normalized}
representation.
For cases where $e = e_{min}$, the mantissa falls within the open interval $(0, 1)$, and
then represent
 \textit{subnormal}
 numbers.\footnotemark
\footnotetext{Both $+0$ and $-0$ are supported, but neither is a subnormal number.
However note that $ulp$ and half of $ulp$ are both normal numbers.}
 We use $ulp$
 as an abbreviation for
 {\em units in 
 the last place} and represents 
 $fp\_value(x)$
 when $s=0,e=0$
 and only the LSB of $m$ is set.
The IEEE standard also details the specifications for 16-bit, 32-bit, and 64-bit floating-point numbers, which are referred as FP16, FP32, and FP64 respectively. With the rise in demand for less but acceptable precision in deep learning, Google introduced the brain-float 16\footnote{\url{https://en.wikipedia.org/wiki/Bfloat16_floating-point_format}} format or BF16. Additionally, NVIDIA unveiled a custom format, notably TensorFloat32\footnote{\url{https://blogs.nvidia.com/blog/2020/05/14/tensorfloat-32-precision-format/}}, tailored for matrix multiplication. This format optimizes for both precision and range, specifically for their  tensor cores. 

\paragraph{Behavioral Portability Issues due to Subnormals}

Our feature-targeted tests include testing for
subnormal support.
To motivate reasons
for such tests in a general context,
consider
two floating-point 
{\em normal} values $a$ and $b$  are close together but not individually equal to $0$.
Suppose we now have an expression $E_1=c/(a-b)$
where $E_1$ is some expression and $c$ is a
normal number.
If the result of $(a-b)$ is a subnormal number
as per an infinite-precision calculation 
but the hardware does not provide support
for subnormals, then the hardware turns
the denominator into $0$
 causing a 
 division-by-zero 
 exception.\footnote{Assuming that pertinent compiler flags are applied.}

 The first problem posed by this situation
 is that some GPUs do not have hardware
 traps for exceptions (~\cite{nvidia-exception-support} confirms this for NVIDIA).
 % \todo[inline]{ilaguna: I wouldn't say many GPUs, but some GPUs.} I will put your comment in the chat. 
 %
 Now if we have another expression $E_2$ 
 similar to $E_1$ and we have $E_1/E_2$; then
 the resulting $\infty/\infty$ results 
 in a NaN (``not a number'') exception---for
 which also GPUs lack adequate 
 exception-trapping support in 
 hardware.
Also note that AMD provides some support for exception trapping~\cite{amd-instruction}, and thus porting from AMD to NVIDIA will turn the lack of subnormal support into
a significant exception-behavior difference.
Thus, a subnormal-targeted test (which we provide) can meaningfully distinguish between GPU behaviors.

\paragraph{Rounding Mechanisms in Floating-Point Arithmetic}
Given the constraints of the floating-point format, rounding becomes essential when a value surpasses its bounds. The IEEE prescribes that rounding should emulate an intermediate result that is infinitely precise and possesses an unbounded range (this is called {\em correct rounding}).
To realize this ideal, supplementary bits (guard or G, rounding or R, and sticky or S, collectively called ``extra bits'', see 
Table~\ref{tab:grs-bits}) are incorporated in the design of IEEE-compliant hardware, with G, R, and S having lower significance (in that order) than the mantissa {\bf least significant bit} (LSB).
These bits are set when operation results are normalized via a right shift (see below for an illustration).
Higham~\cite{higham2002accuracy} notes that a single additional bit will not consistently yield the same outcome as obtaining the precise result followed by rounding. However, incorporating a second guard bit and a third sticky bit (which is the logical OR of all bits that are shifted through the S position) permits correct rounding.
To realize correct rounding~\cite{higham2002accuracy}, there are two requirements: (1) employ three extra precision bits, and (2) employ round-to nearest with ties to even.

\begin{table}[h]
\caption{Rounding Rules of FP Arithmetic. To read this table, first locate the GRS bits. Then decide the result sign and the rounding mode desired. Finally, for all but truncate, add the specified bit to the mantissa least significant bit (LSB). For truncate, set the LSB as per this value.}
\resizebox{\linewidth}{!}{
\begin{tabular}{c|c|cccc}
\hline
\multirow{2}{*}{\begin{tabular}[c]{@{}c@{}}\shortstack{The three \\ extra bits\\ GRS where\\ $(x\vee y)=1$}\\ \end{tabular}}             &  {\shortstack{Result\\
sign}} & \multicolumn{4}{c}{\shortstack{New value for mantissa LSB (add this bit to $m$'s\\  LSB
except for truncate it is assigned to LSB} }                                                                                                                                                                                                                                                                 \\ \cline{3-6} 
                                                                                        &                       & \multicolumn{1}{c|}{\begin{tabular}[c]{@{}c@{}}Round \\up \\ (tow. \\ $+\infty$)\end{tabular}} & \multicolumn{1}{c|}{\begin{tabular}[c]{@{}c@{}}Round \\down\\ (tow.\\ $-\infty)$\end{tabular}} & \multicolumn{1}{c|}{\begin{tabular}[c]{@{}c@{}} RTN-TE\end{tabular}} & \begin{tabular}[c]{@{}c@{}}\shortstack{Round \\to zero\\ (truncate)}\end{tabular} \\ \hline
\multirow{2}{*}{\begin{tabular}[c]{@{}c@{}}0xy\end{tabular}}  & +                     & \multicolumn{1}{c|}{1}                                                     & \multicolumn{1}{c|}{0}                                                       & \multicolumn{1}{c|}{0}                                                                                 & 0                                                          \\
                                                                                        & -                     & \multicolumn{1}{c|}{0}                                                     & \multicolumn{1}{c|}{1}                                                       & \multicolumn{1}{c|}{0}                                                                                 & 0                                                          \\ \hline
{100}    & +  & 
   \multicolumn{1}{c|}{1}   & \multicolumn{1}{c|}{0}   & \multicolumn{1}{c|}{$1$} & 0                           \\

    & -                     & \multicolumn{1}{c|}{0}                                                     & \multicolumn{1}{c|}{1}                                                       & \multicolumn{1}{c|}{$1$}                                                                                 & 0                    \\ \hline
\multirow{2}{*}{\begin{tabular}[c]{@{}c@{}}1xy\end{tabular}} & +                     & \multicolumn{1}{c|}{1}                                                     & \multicolumn{1}{c|}{0}                                                       & \multicolumn{1}{c|}{1}                                                                                 & 0                                                          \\
                                                                                        & -                     & \multicolumn{1}{c|}{0}                                                     & \multicolumn{1}{c|}{1}                                                       & \multicolumn{1}{c|}{1}                                                                                 & 0                                                          \\ \hline
\end{tabular}
}
  %  \vspace{0.5em} % Creates a bit of space between the table and the footnote
\label{tab:grs-bits}
\end{table}
 
\noindent{\bf Description of Rounding} 
For clarity, let us walk
through
a simple
example
which will help read the rest
of this paper with more
assurance:
\begin{compactitem}
    \item 
    {\sl Align:\/}
    (If necessary), make the exponent of the two numbers to be added the same by right-shifting the number with the smaller exponent.
    
    \item {\sl Operate,
    normalize, set extra bits:\/}
    Perform the addition, and then normalize the result; specifically, if the result mantissa is 2 or more in value, bring it within $[1,2)$ by right-shifting the mantissa,  suitably 
    adjusting the exponent.
    This right shift sets through and sets the extra bits.

\item {\sl Round  as per rules, normalize again if needed:\/} Consult Table~\ref{tab:grs-bits} to round or truncate.

\end{compactitem}

%
% Suppose after the operation we have the GRS bits being
% $100$ or $1xy$ where
% $x$ or $y$ is 1.
% %
% This means that the result has a ``half $ulp$ or more extra.''
% %
% Now, if the result is positive and if we are rounding up or
% performing RTN-TE, then
% add one $ulp$ to the result (doing another normalization if the mantissa becomes $\ge 2$).
% If we are doing truncation, we {\em set} the mantissa LSB to 0. All other cases can be read similarly.
 
% %

\noindent{\bf An Example} Consider an FP scheme with one bit mantissa and suppose the result after calculation is positive $1.1100$ in binary or 1.75 in decimal ($GRS=100$ is attached at the end), and let $e=0$.
This cannot be represented using one mantissa bit, and so we must round.
For RTN-TE\footnote{RTN-TE stands for "Round to Nearest, Ties to Even," which is a rounding method commonly used in floating-point arithmetic. When a number falls exactly halfway between two possible rounded values, this method rounds the number to the nearest even value. }
%\bfcmdside{Need to explain RTN-TE?}, we will be adding 1 to 
the mantissa LSB   resulting 
in $10.0100$.
This needs normalization, and after that, the
 result is $1.0010$ 
(and $e=1$)---i.e., 2 in decimal. 
The answer for truncate is 1.

% Building on this, the IEEE standard recommends \textit{rounding to the nearest} as the default mode.~\cite{ieee2019ieee}.
%

\paragraph{Fused Multiply-Add (FMA) Operation}
In contemporary computational architectures, certain machines incorporate hardware components specifically designed to facilitate the Fused Multiply-Add (FMA) operation. As per the IEEE 754 standards, this operation computes $c+(a\cdot b)$ by ensuring two pivotal conditions: (1) computation is performed as though it has infinite precision and an unbounded range, and (2) rounding is applied only once, after the completion of both `*' and `+'. These  are referred to as   FMA conventions in this paper. Matrix multiplication, represented as $A\cdot B + C$, can be conceptualized as a series of blocked Multiply-Add Operations. This operation is natively supported by GPU architectures from both AMD and NVIDIA, as elaborated in Section \ref{sec:gpu-matrix-back}. Given this context, we hypothesize these matrix accelerators  also adhere to these FMA conventions.

\subsection{Matrix Acceleration} 
\label{sec:gpu-matrix-back}
%  In NVIDIA GPUs,
% CUDA cores carry out the basic
% arithmetic operations,
% while
% special function units 
% (now called multiple function units or MUFU)
% perform special math functions,
% and
% finally tensor cores perform matrix operations.
% %
% In AMD GPUs, matrix operations are carried out
% by matrix cores.
% %
% We use the term
% {\em matrix accelerators}
% when we want to cover both GPUs.

    % \item Basic arithmetical operations (Addition, multiplication): CUDA cores/(investigate the corresponding amd cores) arithmetical units. 
    % \item Some special functions: multiple-function-units(MUFU) in NVIDIA GPUs and (?AMD). Also dependent on the software implementation.
    % \item Matrix Multiplication: tensor cores/matrix cores, also depedent on the software implementation

%
Implementing matrix operations efficiently benefits a plethora of numerical algorithms underlying  HPC
and ML~\cite{jack-royal}.
Acknowledging this,
NVIDIA and AMD have developed specialized compute units. NVIDIA's Tensor Cores and AMD's Matrix Cores are designed to optimize matrix operations, enhancing computational speed and efficiency.
 We use the neutral term {\bf matrix accelerator} when referring to either.
Matrix multiplication, represented by the equation $D = A\cdot B + C$, is a foundational primitive in
Linear Algebra (it is a BLAS level 3 operation). 
%
%
% Here, the precision of the 
% $D$ matrix {\em is allowed to be the
% same or lower than that 
% of the input matrices.}\ggcmtside{XL plz fix!}
Equation~\ref{eq:gen-blas-eqn}
for all $i,j$ in the allowed range of matrix indices $1\dots Size$ governs the behavior of
matrix accelerators:
\begin{equation}
\label{eq:gen-blas-eqn}
d_{ij} = a_{i1}*b_{j1}+a_{i2}b_{2j}+...+a_{in}b_{nj} + c_{ij}.
\end{equation}

\subsection{Block FMA}

Existing public documentation on matrix accelerators~\cite{blanchard:hal-02491076,jack-royal} indicates that
they employ the so-called
{\em block FMA} where the calculation in
Equation~\ref{eq:gen-blas-eqn}
is achieved in parallel,
essentially suffering rounding error comparable to doing one scalar FMA.
In other words, if one unrolls 
Equation~\ref{eq:gen-blas-eqn}
into a serial loop and determines the rounding error, then it is clear that each add-multiply step can incur a half $ulp$ error in RTN-TE, thus making the worst-case error grow with $Size$.
This is avoided in block-FMA; we assume block-FMA in the rest of
this paper.

\subsection{Coding Matrix Acceleration}

There are mainly two ways to invoke matrix accelerators:
%---
\begin{enumerate}
    \item  {\em Via High-Level APIs:} One can make use of high-level C++ APIs such as \texttt{nvcuda::wmma} for NVIDIA and \texttt{nrocmwmma:wmma} for AMD. These APIs provide a structured and relatively user-friendly interface to interact with Tensor and Matrix cores, respectively.

    \item {\em Intermediate-Level Assembly Manipulation:} For those delving deeper into the architecture, direct interaction with matrix accelerators unit is feasible through specific instruction sets. Within the NVIDIA platform, this is achieved by the PTX instruction set \texttt{wmma} operations.  In contrast, AMD offers compiler intrinsic instructions such as \texttt{\_\_builtin\_amdgcn\_mfma\_} are tailored for matrix operations.
\end{enumerate}

% We now present a few pragmatic aspects of the ``specification discovery'' exercises presented in this paper.
% %
% First, we assume certain ``obvious symmetry'' with respect to hardware.
%
% For instance, it is difficult to imagine making a piece of hardware fail when adding (say) 7 and 3, or choose a new rounding mode specific to that addition.
%
% Second, we assume that partial products $a_{ip} b_{pj}$
% and
% $a_{iq} b_{qj}$
% (for $p\neq q$)
% behave identically with respect to the accumulation steps of all the additions when all other partial products are zero. 
%
% \footnotetext{FIX FOOTNOTE NUMBER Else there is a fatal error; such outcomes are monitored by our tests but not expected   in the post Pentium FDIV-bug~\cite{pentium} era.}
% \footnotetext{We do separately check for the reduction order---the order in which the additions are carried out---as to whether that order can be externally controlled or not.}

In our work, we employed the high-level API directly, abiding
by all the requirements
for its invocation such as meeting dimensionality restrictions published by manufacturers.
To double-check that matrix accelerator units will be
active during operation, we check the underlying code. For NVIDIA, the presence of \texttt{HMMA/DMMA} in the SASS code indicates that 
the Tensor Cores will be  activated.
For AMD, spotting \texttt{MFMA} in the LLVM intermediate representation 
indicates
the
use of their Matrix Computing Units~\footnote{NVIDIA's official documentation highlights the roles of \texttt{HMMA} and \texttt{DMMA} operations in Tensor Core operations~\cite{nvidia-instruction}
%, found at \url{<NVIDIA_URL>}. 
Similarly, AMD describes the role of \texttt{MFMA} in their official documentation~\cite{amd-instruction}.
These documents also mention the conditions to be met before these units are activated. Leaving nothing to chance,  we check for these instructions explicitly. 
%paper at \url{<AMD_URL>}
}.

% \subsection{Testing Approaches}

% \subsubsection{Feature-Oriented Tests}

% \begin{itemize}
%     \item purpose : reveal undoc features  -
%     e.g. does it keep extra bits?
%     \item TYpes of tests 
% \end{itemize}

% \subsubsection{Random Tests}

% CSmith (first tool)

% Varity (inspired by CSmith)

% \section{Basic Computational Unit Testing}
% \label{sec:function-testing}
% \input{sec3-basic-methodology}

\section{Numerical Behaviors of Matrix Accelerators}
\label{sec:matrix-accel-ma}
\begin{figure}
    \centering
    \includegraphics[scale=.12]{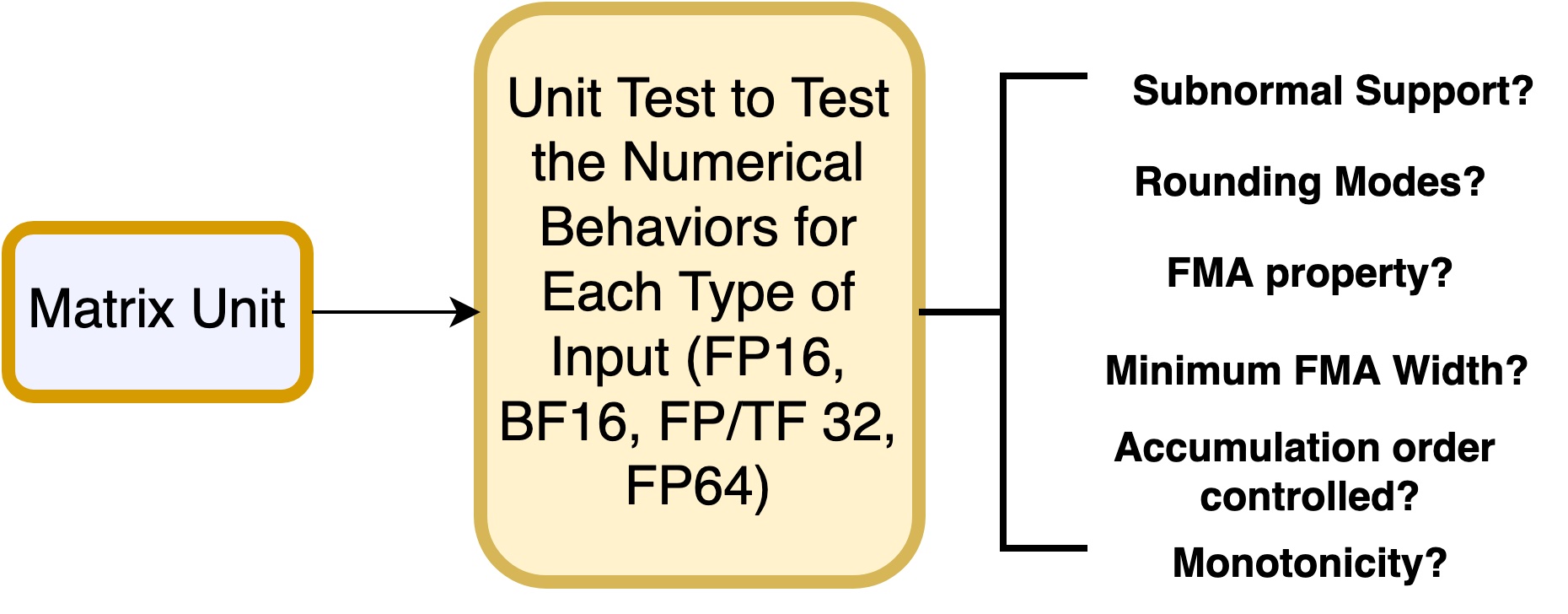}
    \caption{Matrix  Unit Testing Approach. For the matrix accelerator under test, all the properties shown on
    the right are checked (all but monotonicity) or implied (monotonicity)
    by our tests.}
    \label{fig:matrix-core-test}
\end{figure}
 
Matrix computing units
are now an indispensable part of GPU usage in machine learning while also attacting considerable interest from HPC developers~\cite{jack-royal}.
This work aims to close
significant gaps in the official documentation  of NVIDIA and AMD detailing their numerical behaviors by designing tests that highlight specific differences.
%
% Building on the work of Fasi et al.~\cite{higham-tensor-cores} who rigorously tested NVIDIA's tensor cores, 
%
Our overall testing plan is illustrated in Figure~\ref{fig:matrix-core-test}, and detailed in subquent sections.
Given our coverage of close to a dozen high-level features, this section will be hierarchically organized where sections detailing specific tests may be skipped on first reading.

\subsection{High-Level Testing Plan}
\label{sec:hl-testing-plan}

The overall goal of a matrix accelerator is to efficiently support the calculations in producing the $D$ matrix where $D=AB+C$, with $A,B$ and
$C$ also being matrices.
Since all $D$ entries are calculated in an
identical manner, it suffices to focus on
how one particular entry, namely $d_{11}$ is
calculated:

\begin{equation}\label{eq:d11eqn}
d_{11} = a_{11}b_{11}+a_{12}b_{21}+...+a_{1n}b_{n1} + c_{11}
\end{equation}

\begin{figure*}[htb]
    \centering
    \includegraphics[scale=0.08]{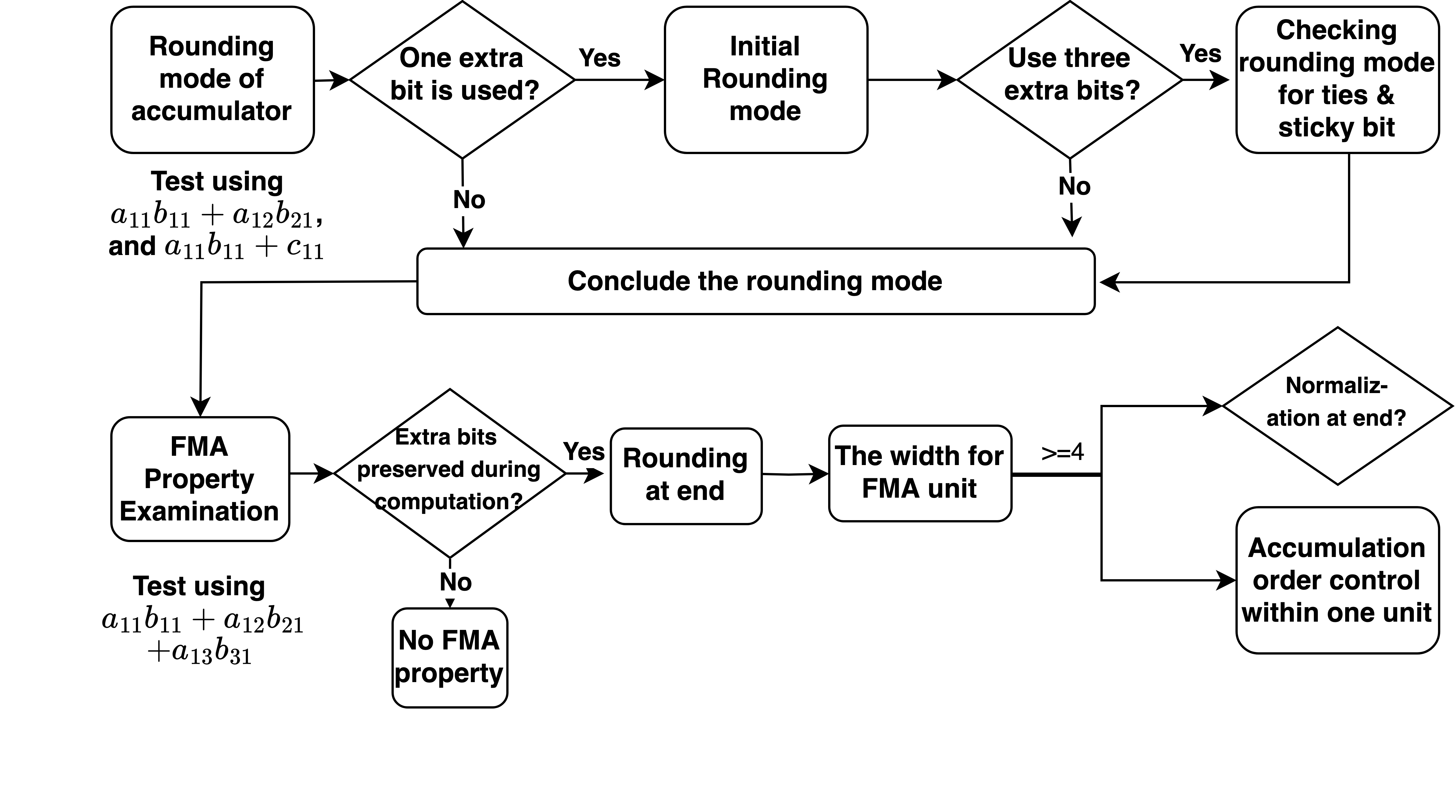}
    \caption{Testing workflow that sharpens each later test based on the
    previous
    ones. First settle the rounding mode of the accumulator
    (T\_rnd\_dir). Then settle the presence of an extra bit; if so then determine the initial
    rounding mode; then settle the use of 3 extra bits (T\_1\_bit, T\_rnd\_dir,
    and
    T\_3\_bits\_fin\_rnd); if so, check for ties and sticky bit. Having concluded the rounding mode, switch to settling FMA properties. Then the extra bits preserved. At that time, we can determine the block FMA width, accumulation order
    control (T\_blk\_fma\_width, T\_acc\_order), and settle whether normalization happens once.}
    \label{fig:pipeline-of-tests}
\end{figure*}

We now focus on each aspect of Equation~\ref{eq:d11eqn}, discuss tests that
target the discovery of the details hidden behind the following features (and these details are the ones that will help contrast various GPUs):
The following tests will now be detailed in their own sections.
{\em An important contribution we make}
is to orchestrate these tests according to the order in the flow-chart in
Figure~\ref{fig:pipeline-of-tests} so that some cases are eliminated or concluded early, allowing later tests to discriminate cases without ambiguity.\footnote{{\bf It is important} to reiterate that the features discovered by these tests are largely
undocumented or hard to find.
Our tests provide a ``one-stop shopping'' experience for quickly determining them {\em and} 
corroborating with documentation (that may change over time.)}
\begin{compactdesc}
% old T2
\item[T\_si\_no:] ``{\sl subnormal in; normal out;}'' i.e., if a computation unit is fed subnormal inputs, can it handle it at the input (without zeroing it), and
produce a normal output?

% old T3
\item[T\_ni\_so:] ``{\sl normal in; subnormal out;}'' i.e., if a computation unit is provided normal inputs, and the computation  resulting in a 
subnormal, then can this subnormal be  output (or will it get zeroed)?

% old T4
\item[T\_sa:] ``subnormal accumulation ok;'' i.e., if a set of subnormals are being added, is the accumulation successful (or is the output getting zeroed)?

% old T5
\item[T\_1\_bit:] ``{\sl at least one extra bit;}'' i.e., is there at
least one extra precision bit in the
computation unit?

% old T6
\item[T\_rnd\_dir:] ``{\sl rounding direction;}'' i.e., determine the rounding direction based on the test outcome: possible test outcomes are to say whether to zero (truncate), down (to $-\infty$), RTN-TE, or up (to $+\infty$) are happening (Figure~\ref{fig:rounding}).

% old T7
\item[T\_3\_bits\_fin\_rnd:] ``{\sl three extra bits are provided, final rounding;}'' i.e., tests that locate if three extra precision bits are provided. It also determines the final rounding direction followed.

\item[T\_prod:] ``{\sl product rounding direction;}'' similar to $T\_rnd\_dir$
except for the product terms during block FMA. 
%We skip presenting the details that  are similar to (Figure~\ref{fig:rounding}).
% \xlcmt{Here, we need to explain that we don't test product for tf32, fp16, bf16 but only test for fp32 and fp64: The multiplication of two floating-point numbers, each having an n-bit mantissa, can potentially yield a product spanning up to 2n bits at its maximum. Taking this into consideration, inputs formats such as fp16, tf32, and bf16—which respectively possess 10, 10, and 7-bit mantissas—don’t experience precision loss when operating within an fp32 environment equipped with a 23-bit mantissa. Contrasting, for the FP32 and FP64 formats, rounding mode determination should be evaluated since the resultant precision can surpasses their representational capacity. I created a test for it, shall we use?To determine this, we can employ the same methodology used for rounding mode assessment during accumulation. The crux lies in devising a test that produces a value beyond the representational capacity of the testing formats. Specifically, for the product a 11 ·b 11 , if we set one term to be 1+2 2 ·ulp+ulp and the other as 1 + 2 −3 , the exact result is 1 + 2 −3 + ulp + 2 −1 · ulp + 2 −2 · ulp + 2 −3 · ulp, with a 111 suffix to the end of the mantissa bit. By integrating negative test scenarios and referencing the rounding tests as depicted in Figure 5, we can deduce the rounding mode used in the multiplication operation. Adopting a similar approach, the testing for three extra bits can be seamlessly conducted.}

% old T8
\item[T\_blk\_fma\_width:] ``{\sl block FMA width;}'' i.e., what is the unit-width for the block FMA operations?

\item[T\_pres\_extra\_acc:] ``{\sl preservation of extra bit during accumulation;}'' 
i.e., are the extra bits preserved during the accumulation of a block FMA unit.

\item[T\_acc\_order:] ``{\sl accumulation order control;}'' i.e., 
can we determine the accumulation order being followed by the block FMA 
during its accumulation stage?

\end{compactdesc}

\subsection{Details of Each Test}

\begin{figure}
    \centering
    \includegraphics[scale=0.078]{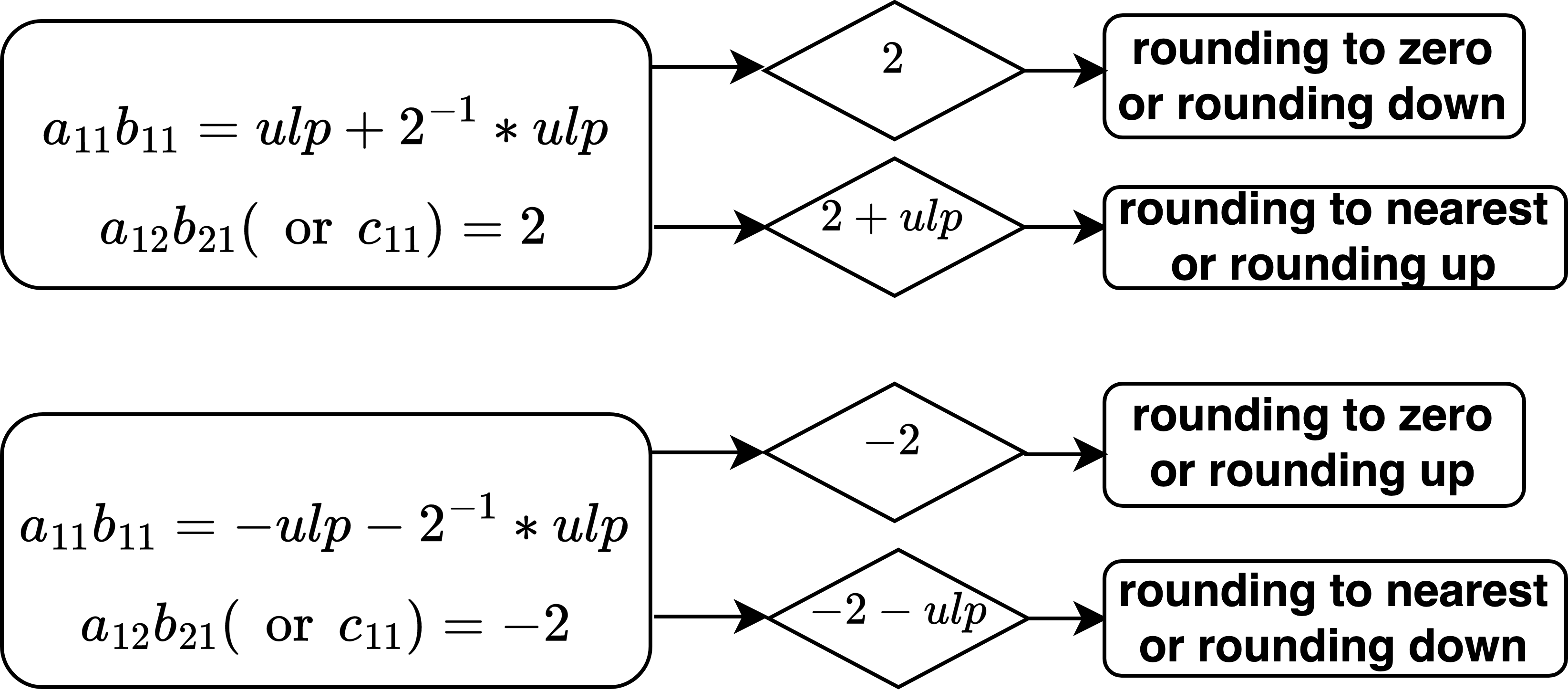}
    \caption{The logic for test T\_rnd\_dir are
    presented here.
    By setting
    the $a_{11}b_{11}$ product 
    as well as
    the 
    $a_{12}b_{21}$ product
    (alternatively the
    $c_{11}$ 
    value)
    to the indicated
    value, the
    execution is carried out
    (all other inputs not mentioned are set to $0$).
    Then
    by 
    examining
    the $d_{11}$
    output,
    we can decide which case we fall into with respect to 
    the rounding
    being used.
    A similar
    logic
    also underlies the $T_{prod}$
    test.}
    \label{fig:rounding}
\end{figure}

 \paragraph{T\_si\_no, T\_ni\_so, and T\_sa:}  

The objective here is to discern whether the matrix unit can handle subnormal numbers, both as input and output.
The tests below will assign specific values to the right-hand side of 
Equation~\ref{eq:d11eqn},
assume full IEEE-compatible subnormal support, and check for 
the expected results under this assumption.
The three tests are now detailed:

 \vspace{.5ex}
     \noindent{\sl T\_si\_no:} 
Initialize $a_{11}$ 
    with an arbitrary subnormal number
    while ensuring that the product $a_{11}b_{11}$ yields a normal number; set
    all other input words
    in the $d_{11}$ equation
    to $0$.  
    Now check
    whether   $d_{11}$ equals $a_{11}b_{11}$; if so, the check passes; else, $d_{11}$ is expected to emerge
    as  zero, when the
    check fails.

 \vspace{.5ex}
\noindent{\sl T\_ni\_so:}
 Initialize $a_{11}$ and $b_{11}$ 
    with 
    arbitrary
    normal numbers
    while ensuring that the product $a_{11}b_{11}$ is a subnormal number. 
    Now examine 
    whether $d_{11}$ is a subnormal value (``pass'') or emerges as zero (``fail''). 

\vspace{.5ex}
\noindent{\sl T\_sa:}
Assign 
    an arbitrary subnormal to $c_{11}$
    while keeping all other inputs at zero.%\ggcmtside{check}
    The test  observes whether $d_{11}$
    is this subnormal (``pass'') or $0$ (``fail'').%\ggcmtside{check}

\begin{figure*}
    \centering
    \includegraphics[scale=0.065]{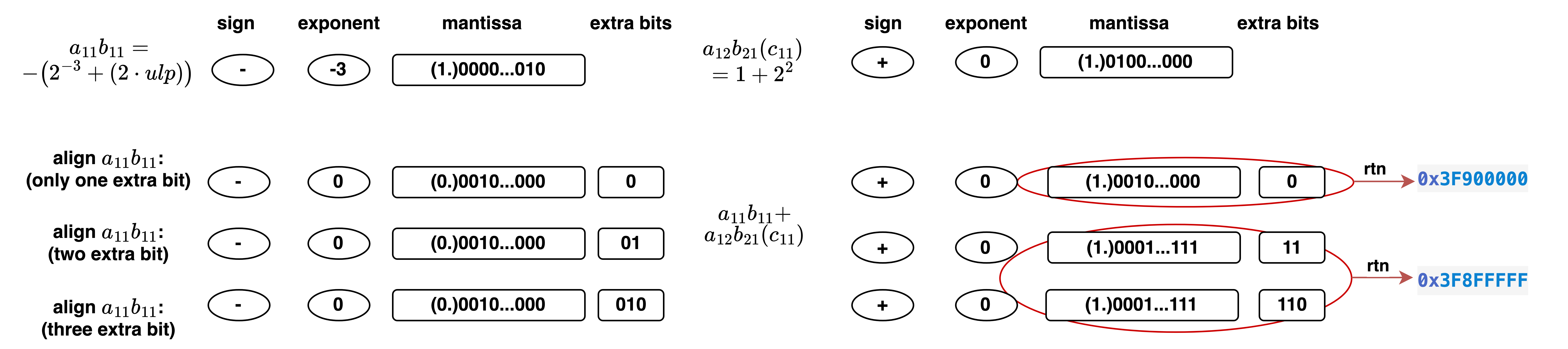}
    \caption{Binary Computation for Two Numbers Addition with Rounding to Nearest Mode. Here is
    how to read this figure. On the left, the
    situation of $a_{11}b_{11}$ (augend) with a specific
    input is shown.  This value
    is aligned since the addend
    ($a_{12}b_{21}$ or $c_{11}$)
    has the higher exponent.
    Alignment under
     one, two, or three extra bits is shown
     underneath $a_{11}b_{11}$.
     The result produced by ``rtn'' (RTN-TE)
     is shown as emitted by the bottom red oval.
    }
    \label{fig:3-extra-rtn}
\end{figure*}

\begin{figure*}
    \centering
    \includegraphics[scale=0.065]{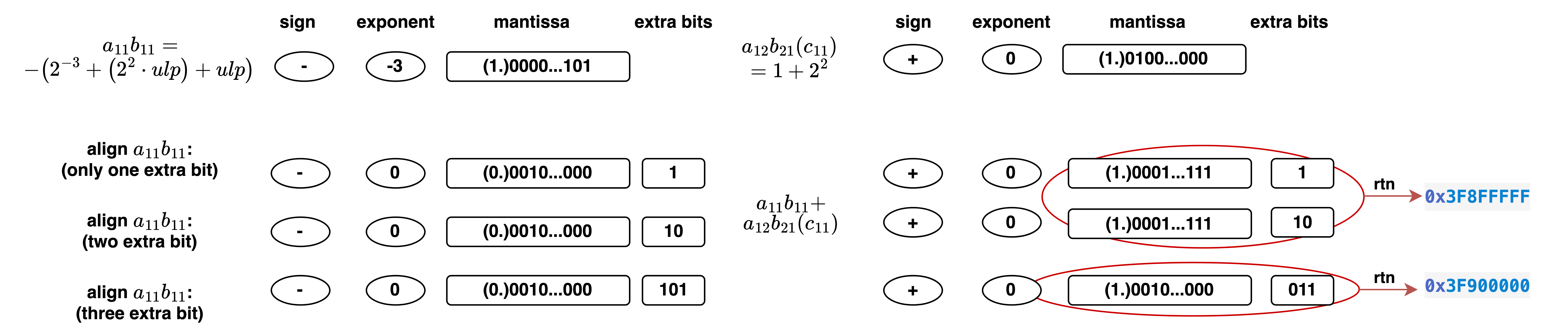}
    \caption{Binary Computation for Two Numbers Addition with Rounding to Zero Mode. Follow the reading suggestions as with Figure~\ref{fig:3-extra-rtn}.}
    \label{fig:3-extra-rtz}
\end{figure*}

% old T5
\paragraph{T\_1\_bit, T\_rnd\_dir, and T\_3\_bits\_fin\_rnd:}
These tests follow the
logic in 
Figure~\ref{fig:rounding} for the first two tests, and
Figure~\ref{fig:3-extra-rtn}
for the RTN-TE case
and 
Figure~\ref{fig:3-extra-rtz}
for the round to zero case.
The tests are now detailed.
Note that the tests can set 
either $a_{12}b_{21}$
or $c_{11}$ to $2$.

\vspace{.5ex}
\noindent{\sl T\_1\_bit:}
We check the result of the operation
$1 - (2^{-1}\times ulp)$  to check
if at least one extra bit is provided.
Aligning $2^{-1} \times ulp$ to 1 in its binary representation necessitates a shift 
amount
equivalent to the mantissa bit length plus one. 
Consequently, if the resultant value remains 
$1 - (2^{-1}\times ulp)$, it implies the existence of an extra bit in computations.

\vspace{.5ex}
\noindent{\sl T\_rnd\_dir:}
Upon confirming the presence of an extra bit, the accumulator's rounding behavior
is assessed
(Figure~\ref{fig:rounding}) .

\vspace{.5ex}
\noindent{\sl T\_3\_bits\_fin\_rnd:}
%--
We finally proceed to
determine if 
  three extra bits are
  present,
  and also determine the
  final rounding modes supported.
  Its logic and implementation
  are now discussed.\footnote{Details provided for the sake of completeness; reading can be postponed.}
 The key aspect of our testing approach was the alignment of three bits during the accumulation process. The goal was to ascertain the effects of preserving one, two, or all three extra bits on the mantissa component. This nuanced behavior was attained via subtraction operations, the intricacies of which are detailed in 
Figures~\ref{fig:3-extra-rtn}
and \ref{fig:3-extra-rtz}.

  Building upon our preliminary understanding of the rounding direction, we embarked on a series of rigorous tests. The essence of these tests is encapsulated in Figures~\ref{fig:3-extra-rtn} and~\ref{fig:3-extra-rtz}, which respectively illustrate the methodologies for rounding-to-nearest and rounding-to-zero modes.\footnote{We highlight these two modes given their adoption in matrix accelerators.}

\paragraph{T\_pres\_extra\_acc:} 
The fact that the extra bits are retained during block-FMA 
accumulation can be confirmed using the expression $1+2^{-1}\cdot ulp + 2^{-1}\cdot ulp + 2^{-1}\cdot ulp$
(by making the accumulation of a block-FMA perform this calculation).%\ggcmtside{check! else it is not clear when this expression is executed.}
If intermediate 
accumulation
steps maintain these extra bits, the ultimate result will be $1+ulp$; else it will emerge as $1$.

\paragraph{T\_acc\_ord:}
  The significance of accumulation order control primarily arises in scenarios employing rounding to zero with the preservation of just one extra bit. Contrarily, when three extra bits are enabled
  (which facilitates a sticky bit with the rounding-to-nearest mode),
  the results remain consistent irrespective of the accumulation order. 
  If only one extra bit is maintained, we must test to ascertain
  the accumulation order. For this,
  one can test {\em all permutations} of the terms
  in the equation $1+2^{-2}\cdot ulp+2^{-2}\cdot ulp+2^{-2}\cdot ulp+2^{-2}\cdot ulp$. Given that only one extra bit is retained in the rounding to zero case, the precision associated with the terms $2^{-2}\cdot ulp+2^{-2}\cdot ulp+2^{-2}\cdot ulp+2^{-2}\cdot ulp$ tends to be lost, {\bf save for when it is computed first}. 
  That is, if we allow all the ``small values'' to add up first, then even with
  one extra bit, we will get the answer $1+ulp$.
  In other words,
  ff we can {\em externally control the order of reduction}
  by assigning these terms to specific positions within an FMA unit, there exists a output yields $1+ulp$ and hence the reduction order is under user control; else not.
  
 %==
   \begin{algorithm}
\DontPrintSemicolon
\caption{Test Minimum Unit for FMA property preservation. The idea is to assign a moving position the $2^{-1}ulp$ value
and when that position goes beyond the width of the block FMA, we get a $1$ output. That index is the FMA block width.}
\label{alg:min-fma-unit}

\KwData{Matrices $a$, $b$, $c$, and $d$. $a's$ row's length $K$.}

Initialize all values in $a$, $b$, $c$ to 0\;
$c_{11} \leftarrow 1.$\;
$a_{11} \times b_{11} \leftarrow 2^{-1} \times \text{ulp}$\;

\For{$i \leftarrow 1$ \KwTo $K$}{
    \If{$i > 1$}{
        $a_{1(i-1)} \times b_{(i-1)1} \leftarrow 0$\;
        % $c_{11} \leftarrow 1.$\;
    }
    $(a_{1i} \times b_{i1}) \leftarrow (2^{-1} \times \text{ulp})$ \\
    Call \textbf{wmma}($a$, $b$, $c$, $d$)\;
    \If{$d_{11} = 1.$}{ \label{line:precision-loss}
       % \tcc{The intermediate result is not preserved, the FMA unit ends.}
        \textbf{break}\;
    }
}

\If{index $<$ K}{ 
    $min\_preserve\_uint \leftarrow index$\;
}
\Else{
    $min\_preserve\_uint$ is larger than $K$\;
    % Call \textbf{printitem}(outfile, "The minimum unit to preserve bits is larger than K.")\;
}

\end{algorithm}

%--
 
% old T8
\paragraph{T\_blk\_fma\_width:}  
To determine the {\em block size} of  the block FMA unit,
we execute a test loop
given in Figure~\ref{alg:min-fma-unit}.
 Within a single FMA unit, precision remains intact throughout computation, and rounding occurs only at the concluding bit position.
 The key idea realized by
 this test is to load-up
 a 1-bit at a pair
 of moving positions
 denoted by
 $a_{1i}$ and $b_{i1}$
 such that
 $a_{1i}\times b\_{i1}$
 is ensured to be 
 half a $ulp$ ($2^{-1}\times ulp$).
  We use the aforementioned equation and shift the last term, $2^{-1}\cdot ulp$, across the matrix multiplication positions as illustrated.
  A loss of precision at Line~\ref{line:precision-loss} (the half $ulp$ vanishes) signals  that
  the initialization occurred
  in a detached FMA unit where it suffers a rounding precision loss;
  the ``end of a FMA unit''
  in effect gets detected
  when the final value is $1$.
 %==
 \paragraph{T\_prod:} 
 Tests for the rounding mode of the product are only performed for FP64 and FP32 inputs. Assuming each multiplier has an $n$-bit mantissa,  their products can occupy only $2n$ bits. Taking this into consideration, the input formats such as FP16, TF32, and BF16---which respectively possess 10, 10, and 7-bit mantissa bits---do not experience precision loss when operating within an FP32 environment where a 23-bit mantissa is used. 

To check product rounding, we can employ the same methodology used for rounding mode assessment during accumulation (Figure~\ref{fig:rounding}). Specifically, for the product $a_{11} \cdot b_{11}$, if we set one term to be $1+2\cdot ulp + ulp$ and the other as $1+2^{-2}$, the exact result is $1+2^{-2} + 3\cdot ulp + 2^{-1}\cdot ulp + 2^{-2}\cdot ulp$, with a \texttt{110} suffix to the end of the mantissa bit. Similarly
we can incorporate negative-number test scenarios
(Figure~\ref{fig:rounding}) and referencing the rounding tests depicted there, we can deduce the rounding mode used in the multiplication operation.


% \input{tables/matrix-unit-numerical}

% \begin{itemize}
%     \item write a plan by Thu eve to finish this -- "this much per day" till say next week Monday
% \end{itemize}

% \section{Related Work}

% \section{Performance Results}
% \input{performance-results.tex}
 
\section{Feature Test Results}
\label{sec:ma-testing-big-table}

\begin{table*}[ht!]
\caption{Results of Analyzing Matrix Accelerators.
Here, ``FMA unit size'' is
the number
of words considered 
before rounding and normalization are
performed
(``Block FMA Size'').
Note that V100 only support FP16.
Further, FP64 is not supported in MI100. 
Last column: "Case 1" is for add/accumulate,
and "Case 2" refers to the product test T\_prod. \cmark is yes,
and \xmark is no.}
\label{fig:matrix-unit-analysis}
\resizebox{\linewidth}{!}{
% \centering{
\begin{tabular}{|l|c|c|c|c|c|c|c|c|}
\hline
Inputs                                                                            & GPU    & \begin{tabular}[c]{@{}c@{}}Subnormal\\ inputs\\handled? \end{tabular} & \begin{tabular}[c]{@{}c@{}}Subnormal\\ outputs\\handled?\end{tabular} & \begin{tabular}[c]{@{}c@{}}Extra bit \\ present?\\How many?\end{tabular} & \begin{tabular}[c]{@{}c@{}}Rounding\\ mode\\exhibited\end{tabular} & \begin{tabular}[c]{@{}c@{}}FMA \\ unit\\ width\end{tabular} & \begin{tabular}[c]{@{}c@{}}Order \\ within one\\  FMA unit \\ is controllable?\end{tabular} & \begin{tabular}[c]{@{}c@{}}Rounding mode for:\\ 1. outputting FP16/BF16\\ (only for FP16/BF16\\  inputs)\\ 2. product (only for \\ FP32/FP64 inputs)\end{tabular} \\ \hline
\multirow{5}{*}{FP16}                                                             & V100   & \cmark                                      & \cmark                                       & 0                                                    & truncate                                                   & 4                                                           & \xmark                                                                        & RTN-TE                                                                                                                                                            \\ \cline{2-9} 
                                                                                  & A100     & \cmark                                      & \cmark                                       & 1                                                    & truncate                                                   & 8                                                           & \xmark                                                                        & RTN-TE                                                                                                                                                            \\ \cline{2-9} 
                                                                                  & H100   & \cmark                                      & \cmark                                       & $\geq 2$                                             & truncate                                                   & $\geq 16$                                                   & \xmark                                                                        & RTN-TE                                                                                                                                                            \\ \cline{2-9} 
                                                                                  & MI100  & \cmark                                      & \cmark                                       & 3                                                    & RTN-TE$^{*}$                         & 4                                                           & \xmark                                                                        & RTN-TE                                                                                                                                                            \\ \cline{2-9} 
                                                                                  & MI250X   & \xmark                                      & \xmark                                       & 3                                                    & RTN-TE                                                     & 1                                                           & N.A.                                                                                         & RTN-TE                                                                                                                                                            \\ \hline
\multirow{4}{*}{BF16}                                                             & A100   & \cmark                                      & \cmark                                       & 1                                                    & truncate                                                   & 8                                                           & \xmark                                                                        & N.A.$^{**}$                                                                                                                                  \\ \cline{2-9} 
                                                                                  & H100   & \cmark                                      & \cmark                                       & $\geq 2$                                             & truncate                                                   & $\geq 16$                                                   & \xmark                                                                        & RTN-TE                                                                                                                                                            \\ \cline{2-9} 
                                                                                  & MI100  & \cmark                                      & \cmark                                       & 3                                                  & RTN-TE                                                     & 2                                                           & \xmark                                                                        & RTN-TE                                                                                                                                                            \\ \cline{2-9} 
                                                                                  & MI250X & \xmark                                      & \xmark                                       & 3                                                  & RTN-TE                                                     & 1                                                           & N.A.                                                                                         & RTN-TE                                                                                                                                                            \\ \hline
\multirow{4}{*}{\begin{tabular}[c]{@{}l@{}}TF32(NVIDIA)\\ FP32(AMD)\end{tabular}} & A100   & \cmark                                      & \cmark                                       & 1                                                  & RTN-TE                                                     & 4                                                           & \xmark                                                                        & N.A.                                                                                                                                                              \\ \cline{2-9} 
                                                                                  & H100   & \cmark                                      & \cmark                                       & $\geq 2$                                             & truncate                                                   & 4                                                           & \xmark                                                                        & N.A.                                                                                                                                                              \\ \cline{2-9} 
                                                                                  & MI100  & \cmark                                      & \cmark                                       & 3                                                  & RTN-TE                                                     & 1                                                           & N.A.                                                                                         & RTN-TE                                                                                                                                                            \\ \cline{2-9} 
                                                                                  & MI250X & \cmark                                      & \cmark                                       & 3                                                  & RTN-TE                                                     & 1                                                           & N.A.                                                                                         & RTN-TE                                                                                                                                                            \\ \hline
\multirow{3}{*}{FP64}                                                             & A100   & \cmark                                      & \cmark                                       & 3                                                    & RTN-TE                                                     & 1                                                           & \xmark                                                                        & RTN-TE                                                                                                                                                            \\ \cline{2-9} 
                                                                                  & H100   & \cmark                                      & \cmark                                       & 3                                                    & RTN-TE                                                     & 1                                                           & \xmark                                                                        & RTN-TE                                                                                                                                                            \\ \cline{2-9} 
                                                                                  & MI250X & \cmark                                      & \cmark                                       & 3                                                    & RTN-TE                                                     & 1                                                           & N.A.                                                                                         & RTN-TE                                                                                                                                                            \\ \hline
\end{tabular}
}
    \vspace{0.5em} % Creates a bit of space between the table and the footnote
    \begin{tabular}{l}
        * RTN-TE = round to nearest and round to even when tie.\\
        ** A100 doesn't support BF16 output. 
    \end{tabular}
\end{table*}

Table~\ref{fig:matrix-unit-analysis} presents our final compilation of results obtained from testing various GPUs. We discuss the results in detail below:

\noindent{\bf Subnormal Supports}
All the GPUs tested support subnormal numbers for inputs and outputs, with the exception of FP16 and BF16 formats of MI250X which does not. It is important to note that the absence of subnormal support could lead to the risk of generating exceptions such as division by zero as mentioned
in \S\ref{sec:bg}.

\noindent{\bf Extra Bits for Computation}
AMD GPUs consistently use three extra bits for precise rounding. In contrast, NVIDIA GPUs have evolved across generations: the V100 does not include any extra bits, the A100 includes one, and the H100 includes at least two extra bits\footnote{Due to our limited access to the H100, we can only test for more than 2 extra bits. We did not conduct further FMA unit width tests for the same reason. We can, however, easily expand our tests to include three extra bits.}. For FP64 inputs, all GPUs incorporate an additional three bits.

\noindent{\bf Rounding Modes}
The chosen rounding mode is consistent across NVIDIA and AMD GPUs, with all models adhering to the chosen mode consistently across generations.

\noindent{\bf FMA Feature}
NVIDIA's V100 has an FMA unit width of 4, and the A100 expands this to 8, as documented. The H100's FMA unit width is suggested to be at least 16, a detail not officially confirmed. For TF32 inputs on NVIDIA GPUs, the FMA unit width is 4, which suits the 19-bit size of TF32. The AMD MI100 maintains FMA features with different widths for FP16 and BF16 inputs, but the MI250X lacks this feature. While FMA units can enhance accuracy, they may complicate the porting of CPU algorithms which do not typically support blocked FMA operations.

\noindent{\bf Rounding Mode for Outputting FP16 and BF16}
We have examined the rounding mode used when GPUs output FP16 and BF16. All GPU models use the RTN-TE rounding mode. We hypothesize that the conversion to lower precision is performed after the computation at full precision.

\noindent{\bf Rounding Mode for Product}
For products involving FP32/FP64 inputs, all GPUs utilize the RTN-TE mode, demonstrating consistency in following IEEE floating-point arithmetic standards.

\section{Exhibiting Porting Danger in a Matrix Multiplication Routine }
\label{sec:new-xl-section}

% We now illustrate the use of our
% tests to cause a result differences.

% \subsection{Decription of Matrix Multiplication Example}

% % basic idea of the solver
% % relevant parts of the 
% % subroutine actually run

% \subsection{Construction of the Extra Bits Test}

% % Logiv behind the test, 
% % and the test itself
% % shown compactly
  
% \subsection{Test Results and Discussions}

% % provide the test results
% % discuss why the difference
% % provide a listing snippet if it helps

We now illustrate an example in which we perform a simple matrix multiplication to demonstrate how these subtle implementation difference in GPU architectures can vary the numerical outcomes.
We analyze the matrix multiplication equation 

$$ D=\alpha \cdot A \cdot B + \beta \cdot C $$

with matrices \(A\) and \(B\) in FP16 format (sized \(m\cdot k\) and \(k\cdot n\), respectively) and matrices \(C\) and \(D\) in FP32 format (both sized \(m\cdot n\)). Here, \(\alpha = -1\) and \(\beta = 1\), and we set the matrix dimensions to \(m=n=k=2^{13}\).

For matrix \(C\), \(C_{ij} = 2^{20}\) for all \(i\) and \(j\). In matrix \(A\), \(A_{i0} = 2^{10}\), \(A_{ij} = 2^{-2}\) for odd \(j\), and \(A_{ij} = 2^{-3}\) for even \(j\) (except \(j=0\)). For matrix \(B\), \(B_{0j} = 2^{10}\), with other \(B_{ij}\) values set at \(2^{-3}\).
In this scenario, each element of matrix \(D\) is calculated as follows (note that all $D_{ij}$ will be the same):
\begin{align*}
 D_{ij} & = -(A_{i0}\cdot B_{0j} + \sum_{j\%2=1} A_{ij}\cdot B_{ij} + \sum_{j\%2=0}^{j\neq 0} A_{ij}\cdot B_{ij}) + C_{ij}\\
        & = -(2^{10} \cdot 2^{10} - \sum_{2^{12}} 2^{-2} \cdot 2^{-3} -  \sum_{2^{12}-1} 2^{-3} \cdot 2^{-3}) + 2^{20}\\
        & = 2^{7} + 2^{6} - 2^{-6} \approx 191.99218 %- GG fixed the + 2^{-6} to - 2^{-6}
\end{align*}

Note that the terms $2^{-2} \cdot 2^{-3}$ and $2^{-3} \cdot 2^{-3}$ require bit shifts (25 bits or 26 bits) to align \(2^{20} = 2^{10} \cdot 2^{10}\). Thus there would be precision loss for FP32 computation unit. This loss may vary depending on the number of extra bits preserved and the length of the FMA operation.
{\bf This variation is what   produces the sharp result-difference that we observed.}

Specifically, we observed these (rather highly different) $D_{ij}$   values computed using a simple GEMM implementation on different GPUs {\em for the very same $A$, $B$, and $C$ matrix inputs}: \(0\) on NVIDIA A100, V100, AMD MI250 and CPU; \(255.875\) on AMD MI100; and \(191.875\) on NVIDIA H100. 

These discrepancies highlight the importance of understanding hardware-specific computational 
feature 
differences; ignoring these and porting across GPUs can
vary results across this wide range.

\vspace{.5ex}
\noindent{\bf Importance
of This Pattern, Consequences:\/}
The pattern \(D=C - A\cdot B\) (where \(\alpha = -1\) and \(\beta =1\)) is closely related to trailing matrix updates \(A_i = A_i - P_iT_i\) used in a mixed-precision GMRES (Generalized Minimal Residual Method) iterative refinement algorithm~\cite{haidar2018harnessing, haidar2020mixed}. This approach is embedded in cuSolvers\footnote{\url{https://docs.nvidia.com/cuda/cusolver/index.html\#cusolverirsrefinement-t}}. In general, a computation of the type \(D=C - A\cdot B\) is part of a standard BLAS (Basic Linear Algebra Subprograms) level 3 family. It is ubiquitous in various numerical linear algebra computations.
%~\cite{duff2002BLAS} % use for BLAS
This observation spawns future directions discussed in \S\ref{sec:conc}.
 
% {\bf trailing matrix update} routine
% separately, 
% exploiting
% our understanding, causing
% a result shift
% from 11,261 to 11,264\xlcmtside{I don't think we can say we tested this routine separately. 11,261 to 11,264 is just a MM I ran for ten iterations. I then found a better way which is the 0, 255 example. I think we don't need to mention 11,261 to 11,264 here. }

 %<=

\section{Concluding Remarks}
\label{sec:conc}

Matrix accelerators are an important emerging component in the
computational landscape, yet very little can be easily determined
about their numerical behavior.
We study five such accelerators in this paper (with many more available for
use, but with even less documented).
One can learn very little by testing these units on random inputs.
We observe that by exploiting our understanding of basic IEEE floating-point
semantics and whatever is published about these units (e.g., that they
perform block FMA), we can devise tests that target
many key attributes such as subnormal support, rounding modes chosen, the
number of hidden bits, accumulation order control, and width of the basic
FMA blocks.
{\bf Manufacturers select these features largely based on the cost of implementation};
for instance, supporting one extra bit is cheaper; and that makes more
demanding rounding modes (e.g., round-to-nearest) impossible to attain.
Yet {\em manufacturers may have targeted these accelerators for their own
set of priority applications}.
For instance, they may have designed a set of features that make machine learning
fast and efficient.
They might view it as
the HPC developer's ``fault'' for using such matrix accelerators to perform
HPC.
On the other hand, HPC programmers are unaware of many of these dangers such
as six orders of magnitude difference in results---unless they are lucky
to
choose such input matrices.
By designing focused feature-targeted tests, we help foresee pitfalls.

Our tests are {\bf not foolproof}.
All we can say is that we assume significant levels of symmetry in design.
For instance, if a manufacturer changes the precision of
the diagonal outputs $d_{11}$, $d_{22}$, ...,
$d_{NN}$ for some reason (say because they determine the Eigenvalues in some cases),
then all bets are off!
Our tests are still valuable in flagging certain porting decisions as dangerous, thus
{\em adding to} the overall porting strategies.

% %
% If our tests assure that many of the ``good features'' are present, we still
% require users to exercise other checks before they fully trust code-porting.

We do have more findings than have been highlighted.
For example, we 
have discerned that the tensor cores operates with a width of 8.
This observation is in agreement
with the details revealed in NVIDIA's white paper~\cite{nvidia2020nvidia} that the tensor cores' dimensions are $8\times 4\times 8$. 
Our tests also reveal that the hardware unit size for AMD GPUs, 
which is not documented as far as we know, is merely $1$.
This can perhaps help explain rounding errors that might be higher
(or might point to a hardware implementation decision taken by AMD).
However, this agrees with what CPUs also follow.
Additionally, the rounding mode employed by the AMD matrix accelerator adheres to the IEEE 754 standard.
This is significant as it suggests {\em a good possibility of producing results 
on AMD Matrix Cores that are 
consistent with those computed on many CPUs.}

Another inference from our tests is that 
 our conclusions on monotonicity match the observations
 in~\cite{mikaitis2023monotonicity} about extra bit 
 requirements.
We  can show that
 NVIDIA Tensor Cores  can violate
 monotonicity, and for AMD GPUs, because of 3 extra bits, monotoncity will be  preserved.

%\subsection{Discussion of Numerical Results}

One exciting direction triggered by the {\em trailing
matrix updates} pattern is
that many other such patterns may exist in one's implementation of linear-algebra routines as well as 
core routines in other
areas.
This suggests developing tests
for patterns  found in other application spaces.
We would also like to work on
  {\em generalizing} our tests  using formal methods~\cite{z3,DBLP:journals/jar/BoldoJLM15}
 which may allow groups to check their tests for
 consistency and overlaps.
%--

% \begin{verbatim}
    
% * other porting issues - to check/discuss

%   - whether exit() by some threads deadlock a syncthread?
%     > NV - no
%     > AMD - ye

% * - whether AMD assumes warpsize = 64 always? (NV does not)
%     > any issues?

% \end{verbatim}

% \section*{Acknowledgment}

\bibliographystyle{unsrt}%{ieeetr} 
 
\bibliography{arxiv-CCGrid-2024-Matrix-Cores}

\end{document}